\documentstyle[twocolumn,aps,prl,overcite]{revtex} 
\input epsf

\tighten

\def\mxth{\mathsurround=0pt }
\def\xversim#1#2{\lower2.pt\vbox{\baselineskip0pt \lineskip-.2pt
    \ialign{$\mxth#1\hfil##\hfil$\crcr#2\crcr\sim\crcr}}}

\begin{document}
 

\title{ Cosmological Tracking Solutions }

\author{ 
Paul J. Steinhardt,$^{1,2}$ Limin Wang$^{1,3}$ and 
Ivaylo Zlatev$^1$ 
}

\address{$^1$Department of Physics and Astronomy,
University of Pennsylvania,
Philadelphia, PA 19104 
 \vspace{.1in} \\
$^2$Department of Physics, Princeton University, Princeton, NJ 08540 \\
$^3$Department of Physics, Columbia University, New York, NY 10027
}

\maketitle

\begin{abstract}

  A substantial fraction of the energy density of the universe
may consist of quintessence in the
form of a slowly-rolling scalar field.  Since the energy density of
the scalar field generally
decreases more slowly than  the
matter energy density, it appears that the ratio of the two densities must
be set to a special, infinitesimal value in the early universe in order
 to have the two densities nearly coincide today.
  Recently, we introduced the notion of tracker fields
   to  avoid this initial conditions problem.
   In the paper, we address the following questions:  What is the general
   condition to have tracker fields? What is the relation between the
   matter energy density and the equation-of-state of the universe imposed
   by tracker solutions? And, can tracker solutions explain why quintessence
   is becoming
   important today rather than during the early
   universe?

\end{abstract} 
\pacs{PACS number(s): 98.80.-k,98.70.Vc,98.65.Dx,98.80.Cq}

\section{Introduction}

Quintessence\cite{Cald98} has been proposed as the missing energy component that must
be added to the baryonic and
matter density in order to reach the critical density.\cite{ostriker,Turner}  
Quintessence is
a dynamical, slowly-evolving, spatially
inhomogeneous component with negative pressure.  An example is a scalar
field $Q$ slowly rolling down
its potential 
$V(Q)$.\cite{Cald98,TurnW,Weiss,Ratra,Wett,FHSW,CDF,Ferreira,Cope}   
For a general scalar field,
the pressure is $p_Q=\frac{1}{2}\dot{Q}^2-V$, and the energy density
is $\rho_Q=\frac{1}{2}\dot{Q}^2 +V$. For a slowly rolling 
scalar field, the pressure can be 
negative if the kinetic energy is less
than the potential energy. For quintessence,
the equation-of-state, defined as 
$w_Q=p_Q/\rho_Q$,
lies between 0 and $-1$.   Depending on $V(Q)$, $w_Q$ may be constant,
slowly-varying, rapidly varying or
oscillatory.\cite{Cald98}

A key problem with the quintessence proposal is explaining why $\rho_Q$
and the matter energy density should
be comparable today. There are two aspects to this problem.
First of all, 
throughout the history of the universe, the two
densities decrease at different rates,
so it appears that the conditions in the early universe have to be set
very carefully in order for the energy
densities to be comparable today.  We refer to this issue of 
initial conditions as the ``coincidence
problem."\cite{CMBdata} This is a generalization of the flatness problem
described by Dicke and Peebles in 1979;\cite{Peebles}
the very same issue arises with a cosmological constant as well.
A second aspect, which we  call the ``fine-tuning problem," 
is that the value of the quintessence energy density
(or vacuum energy or curvature) is very tiny compared to typical
particle physics scales.  The fine-tuning condition is
forced by direct measurements; however, the initial conditions or coincidence
problem depends on the theoretical candidate for the missing energy.

Recently, we introduced a form of quintessence called ``tracker fields"
which avoids  the coincidence
problem.\cite{zlatev1}
Tracker fields have an equation-of-motion with attractor-like
solutions in which a very wide range of
initial conditions rapidly converge to a common, cosmic evolutionary track.
The initial value of $\rho_Q$ can vary by nearly 100 orders of magnitude
without altering the cosmic history.  The acceptable initial conditions include
the natural 
possibility of equipartition  after inflation --
nearly equal energy density in $Q$ as in the other 100-1000 degrees
of freedom ({\it e.g.},  $\Omega_{Qi} \approx 10^{-3}$).
Furthermore, the resulting cosmology has desirable properties.
The equation-of-state $w_Q$ 
varies according to the
background equation-of-state $w_B$.  When the universe is
radiation-dominated ($w_B=1/3$), then
$w_Q$ is  less than or equal to  1/3 and 
$\rho_Q$ decreases less rapidly than the
radiation density.  When the
universe is matter-dominated ($w_B=0$), then $w_Q$ is less than zero and
$\rho_Q$ decreases less 
rapidly than the matter density.  Eventually, $\rho_Q$ surpasses the
matter density and becomes the dominant
component.  At this point, $Q$ slows to a crawl and $w_Q \rightarrow -1$
as $\Omega_Q \rightarrow 1$
and the universe is driven into an accelerating phase.  
These properties seem to match current observations well.\cite{Wang98b}

In some cases, $w_Q$ is very nearly equal to $w_B$ during the radiation-
and matter-dominated epochs, and
some authors have focused on the
fact that the $Q$ energy density decreases at nearly the same
rate as the background energy density.\cite{Ferreira,LidSher}    
However, our focus is
on the fact that the  cosmology is independent of initial
conditions, which includes but does not require $w_Q \approx w_B$
during the matter- and radiation-dominated epochs.
Hence, our use of the term ``tracker" is meant to refer to
solutions joining a common evolutionary track, as opposed to tracking
closely the background energy density.

It is also an interesting point that some tracker solutions do not
 require small mass parameters to obtain a small energy density
  today;\cite{zlatev1,binetruy} whether this is a satisfactory solution to the
   fine tuning problem is debatable, though. We do not explore this 
   issue in this paper.

The tracker solution differs from a classical dynamics attractor because 
it  is  time-dependent:  during tracking,
$\Omega_Q$ increases steadily.  This contrasts with the 
 solutions based on exponential potentials
 recently discussed by Ferreira and 
Joyce\cite{Ferreira} 
in which $\Omega_Q$ is constant during the matter dominated epoch.
For this kind of solution,
$\Omega_Q$ is constrained to be small ($\Omega_Q \le 0.15$)
 at the beginning of matter domination 
in order that large-scale structure be formed, but then it cannot
change thereafter. Hence, it remains a small, subdominant
component.  (See,  however, discussion of undershoot and
overshoot solutions later in this paper for a possible way
around this constraint.)
The tracker solution is more desirable, though, because it enables 
the $Q$-energy to eventually overtake the matter density and induce
a period of accelerated expansion, which produces a cosmology
more consistent with measurements of the matter density, large
scale structure, and  supernovae
observations.\cite{Wang98b,supernova} 

An important consequence of the tracker solutions is the
prediction  of a relation between 
$w_Q$ and $\Omega_Q$ today.\cite{zlatev1}  Because tracker solutions are 
insensitive to initial conditions, both $w_Q$ and $\Omega_Q$
only depend on $V(Q)$.  Hence, for any given $V(Q)$, once $\Omega_Q$
is measured,
$w_Q$ is determined.
In general, the closer
that $\Omega_Q$ is to unity,
the closer $w_Q$ is to $-1$.  However, since $\Omega_m \ge 0.2$ today,
there is a sufficient gap between
$\Omega_Q$ and unity that $w_Q$ cannot be so close to $-1$.   We find that
$w_Q \gtrsim -0.8$ for practical
models.  This $w_Q$-$\Omega_Q$ relation, which
makes the tracker field proposal
distinguishable from the cosmological constant, will be explored
further in this paper.

The purpose of the present paper is to expand on our introductory
article on tracker fields and the
coincidence problem.  In particular, we want to go beyond the specific
examples of tracker potentials
studied before 
and address in a general way the following  questions (the relevant
section is shown in parentheses):
\begin{itemize}
\item What is a tracker solution? (II.B)
\item What potentials $V(Q)$ have tracker solutions and which do not?
(III.A-C,F)
\item How does  convergence of diverse initial 
conditions  to a common track  occur? (III.D)
\item What additional conditions on tracking potentials 
are required to make them useful and practical? (III.H)
\item How does tracking result in the prediction of
an $\Omega_Q$-$w_Q$ relation? (IV)
\item What range of $w_Q$ today  is possible for tracking solutions
according to the $w_Q$-$\Omega_Q$ relation? (IV)
\item Why is the $Q$-field first beginning to dominate at this late
stage of the universe rather than at some early stage?  (V)
\end{itemize}

\section{Definitions and Basic Equations}

We shall consider a scalar field with present equation-of-state
 $-1<w_Q<0$ in a flat
cosmological background (consistent with inflation).  The ratio of the
energy density to the critical density
today is $\Omega_Q$ for the $Q$-field and $\Omega_m$ for the baryonic
and dark matter density where
$\Omega_m + \Omega_Q=1$.  We use dimensionless units where the 
Planck mass is $M_p=1$.

\subsection{The Tracker Equation}

The equation-of-motion for the $Q$-field is
\begin{equation}
\ddot{Q}+ 3H\dot{Q} + V' =0
\end{equation}
where
\begin{equation}
H^2 = \left(\frac{\dot{a}}{a}\right)^2=\kappa (\rho_m + \rho_r + \frac{1}{2} \dot{Q}^2 + V)
\end{equation}
where $a$ is the Robertson-Walker scale factor,
 $\rho_m$ is the matter density, $\rho_r$ is the radiation energy
 density, and
$\kappa=\frac{8 \pi  }{3 }$. 
The definition of the equation-of-state is
\begin{equation} \label{eos}
w_Q = \frac{p}{\rho} = \frac{\frac{1}{2}\dot{Q}^2-
V}{\frac{1}{2}\dot{Q}^2 + V}.
\end{equation}

It is extremely useful to combine these
 relations into an unfamiliar
form, which is the form  we would like the reader to
have in mind when we refer to 
the  ``equation-of-motion":
\begin{equation} \label{teq}
\pm \frac{V'}{V} = 3 \sqrt{ \frac{\kappa}{\Omega_Q}} \sqrt{1+w_Q}
\left[ 1+ \frac{1}{6} \frac{d\, \ln{x}}{d\,\ln{a}}\right]
\end{equation}
where $x=(1+w_Q)/(1-w_Q)= \frac{1}{2}\dot{Q}^2/V$ is the ratio of the
kinetic to potential energy density for $Q$  and prime means 
derivative with respect to $Q$. The $\pm$ sign depends on 
whether $V'>0$ or $V'<0$, respectively. 
The tracking solution (to which general solutions converge)
has the property that $w_Q$ is nearly constant and 
lies between $w_B $ and $-1$.   For $1+w_Q = {\cal O}(1)$,
$\dot{Q}^2 \approx \Omega_Q H^2$ and
the  equation-of-motion, Eq.~(\ref{teq}),
dictates that 
\begin{equation}
\frac{V'}{V} \approx \frac{1}{\sqrt{\Omega_Q}} \approx \frac{H}{\dot{Q}}
\end{equation}
for a tracking solution;
we shall refer to this as the ``tracker condition."

An important function is $\Gamma \equiv V''V/(V')^2$, whose properties
determine whether tracking solutions exist.  Taking the derivative
of the equation-of-motion with respect to $Q$ and combining with the
equation-of-motion itself, we obtain the equation:
\begin{eqnarray} \label{Gamma}
	&& \Gamma \equiv \frac{V''V}{(V')^2}
= 1 + \frac{w_B-w_Q}{2(1+ w_Q)} - 
\frac{1+w_B-2w_Q}{2(1+w_Q)} \frac{\dot{x}}{6+\dot{x}}  \nonumber \\
	&&
-\frac{2}{(1+w_Q)} \frac{\ddot{x}}{(6+\dot{x})^2}
\end{eqnarray}
where
$\dot{x} \equiv d \,\ln{x}/d \, \ln{a}$
and $\ddot{x} \equiv d^2 \,\ln{x}/d \, \ln{a}^2$.
We will refer to this equation as the ``tracker equation."

\subsection{What is a tracker solution?}

We shall use the term ``converging" to refer to the attractor-like
behavior in
which solutions to the equation-of-motion are drawn towards
a common solution, $\tilde{Q}(t)$. 
We shall argue that converging solutions with $w_Q \le w_B$ solve
the coincidence problem.
 A tracker solution is one which 
undergoes long periods of converging.  
The converging period should be long enough
so that any residue of initial condition is washed away and
the solution today only depends on parameters in the potential.
For a potential $V(Q) = M^4 \tilde{v}(Q/M)$ (where $\tilde{v}$ is 
a dimensionless function of $Q/M$), there is a family of tracker
solutions parameterized by $M$.  The value of $M$ is determined by
the measured value of $\Omega_m$ today (assuming a flat universe).

We make the (fine)
distinction between 
``tracking" and ``converging" because
tracker solutions typically go through some periods when solutions
are not approaching one another.
In particular, 
during transitions in the background equation-of-state, such
as the transition from radiation- to matter-domination, there 
are brief intervals during which $w_Q$ changes rapidly
and  general solutions are not drawn towards the tracker solution; 
fortunately,  these
transitory periods are too short to spoil the advantages of
tracking solutions.  

To be complete  in classifying the possibilities,
one should also consider 
potentials in which $Q$ first passes through a regime where
solutions converge and then ends up in  a regime where they do 
not. We refer to these as ``hybrid  models" below.
By the time the second regime is reached, sufficient convergence
of solutions may
have already
occurred to  insure that the desired range of  initial conditions  track along 
a common solution  in the second regime.  

\section{Properties of Tracking Solutions and Tracking Potentials}

In the following section, we prove the key properties of tracker 
solutions.  Each subsection leads with a summary statement 
of the result
for those who wish to skip the detailed mathematical arguments.

\subsection{What potentials  have converging behavior and
tracker solutions and which do not?}

Our  central theorem 
is that tracking behavior with $w_Q < w_B$ occurs for any 
potential in which  $\Gamma \equiv V''V/(V')^2 >1 $ and
is nearly constant ($|d (\Gamma-1)/H d t| \ll |\Gamma -1| $) 
over the range of plausible initial $Q$.
For the cases
of interest, the range extends to $V(Q)$ equal to the initial background 
energy density $\rho_B$ down to $V(Q)$ equal to the background
density at matter-radiation equality, a span of over 100 orders
of magnitude. Although $w_Q < w_B$ is  what is desired for most
practical applications, we also show that
tracking 
behavior  with $w_B<w_Q< (1/2)(1+w_B)$  is possible provided
 $1-(1-w_B)/(6+2 w_B)<\Gamma <1$ and nearly constant. Tracking does not
occur for $\Gamma<1-(1-w_B)/(6+ 2 w_B)$.

Hence, testing for the existence of tracking solutions reduces
to a simple condition on $V(Q)$
 without having to solve the equation-of-motion directly.  
In particular, the condition that  $\Gamma \approx \, {\rm constant}$ can
be evaluated either by testing 
\begin{equation}
\left|\Gamma^{-1} 
\frac{d (\Gamma-1)}{H d t}  \right|  \approx
\left| \frac{\Gamma'}{\Gamma \, (V'/V)} \right|\ll 1 
\end{equation}
over   range
of $Q$ corresponding to the allowed initial conditions;  the middle
expression is easily computed knowing  $V(Q)$ only without having
to solve an equation-of-motion. (Here we have used the tracker 
condition
$\dot{Q}/H \approx \sqrt{\Omega_Q} \approx V'/V$ which applies
for the tracker solution.)
 An equivalent condition is that
 that  $\Delta \Gamma/\Gamma \ll 1 $, \
where $\Delta \Gamma$ is the difference between the maximum and minimum
values of $\Gamma$ over the same range of $Q$.
The condition $\Gamma >1 $  is equivalent to the 
constraint that $|V'/V|$ be decreasing
as $V$ decreases. 
These conditions encompass an extremely broad range of potentials,
including inverse power-law potentials ($V(Q) = M^{4+\alpha}/Q^\alpha$ 
for $\alpha >0$) and combinations of inverse power-law terms ({\it e.g.},
$V(Q) = M^4 {\rm exp}(M/Q)$).
Some potentials of this form are suggested by
particle physics models with
dynamical symmetry breaking or nonperturbative
effects.\cite{binetruy,AC,barreiro,gaillard,barrow,hill,partphys}

If $w_Q< w_B$, converging behavior   does
not occur for potentials in which  $|V'/V|$ strictly increases 
as $V$ decreases ($\Gamma <1$).
This category includes quadratic potentials and most examples of 
quintessence models in the literature.
Instead of converging behavior,
these models 
require specially tuned initial conditions to obtain an acceptable
value of $\Omega_Q$ today, as discussed in Section III.B.

For $(1/2) (1+w_B)> w_Q > w_B$, 
converging behavior 
 does occur if $V'/V$ is strictly increasing
as $V$ decreases  ($\Gamma <1$); 
however, as we explain in Section III.F, potentials 
with $w_Q >w_B$ do not produce viable cosmological models:
 if $w_Q> w_B$ and $\Omega_Q \ge 1/2$ today (as suggested by 
 current observation)
then $\Omega_Q$  must exceed $\Omega_m$ in the past and there is no
  period of matter-domination or structure formation.

Hybrid potentials are possible in which converging behavior only occurs
for a finite period during the early universe,  provided that
the  time is sufficiently 
long to bring together solutions whose initial conditions span the range
of practical interest.  For example, one can construct potentials
in which $|V'/V|$  decreases at first
as $Q$ rolls downhill, and then $|V'/V|$  begins to increase.
  Another possibility is that $|V'/V|$ increases
at first and $w_Q> w_B$, as discussed in the previous paragraph.  We 
pointed out that this condition allows convergence, but
cannot be maintained up to the present
for practical reasons of structure formation. However, the condition could
be maintained for a long, finite period in the early universe so long
as it is terminated before matter-domination.

\subsection{Why do 
 models with increasing $|V'/V|$ and $w_Q< w_B$ 
 fail to solve the coincidence problem?}

For  potentials in which $|V'/V|$ increases as $V$ decreases ($\Gamma <1$),
the LHS of the 
the  equation-of-motion Eq.~(\ref{teq}) is increasing. If $w_Q < w_B$,
 $1/\sqrt{\Omega_Q}$ on the RHS is decreasing as $Q$  rolls downhill.  
Hence, the tracker condition, $|V'/V|\sim 1/\sqrt{\Omega_Q}$, cannot
be maintained and $w_Q$ cannot be maintained at a nearly constant
value (different from $-1$) for any
extended period.

In particular, 
extrapolating backwards in
time, $|V'/V|$ is strictly  decreasing and $1/\sqrt{\Omega_Q}$ is strictly 
increasing. 
The only way to satisfy the  equation-of-motion, Eq.~(\ref{teq}), is
to have $w_Q$ approach $-1$. But, this corresponds to the $Q$-field
freezing at some  value
$Q=Q_i$ after only a few Hubble times. 
The energy density of the frozen field is only slightly higher than
the current 
energy density.  Consequently, to 
obtain this cosmological solution, one has to have special initial
conditions in the early universe (after inflation, say)
that set $Q$ precisely to $Q_i$
and the $Q$-energy density to $\rho_{Q_i}$, a value nearly 100 orders
of magnitude smaller than the background energy density.  It is 
precisely this tuning of initial conditions (the coincidence problem)
which we seek to avoid.

We note that many quintessence models  ($w_Q<0$) 
discussed  in the literature
fall into this non-tracking class and require
extraordinary tuning of initial conditions.
Simple examples include the harmonic potential, $V(Q) = M^2 Q^2$,
and the sinusoidal potential, $V(Q)= M^4 \,[\cos{(Q/f)}+1]$.
In these models, the initial value of $Q$ must be set to a specific
value initially in order to obtain the measured value of $\Omega_Q$
today.

\subsection{What determines the tracker solution?}

Our claim is that solutions converge to a  tracker solution if:
(a) $\Gamma(Q) = V''V/(V')^2$ is nearly constant; and (b)
$\Gamma >1 $ for $w_Q <w_B$ or $\Gamma <1$ for $(1/2)(1+w_B)>w_Q> w_B$.
We have explained in the previous section why $\Gamma$ must be 
greater than unity if $w_Q <w_B$ or else the tracking 
condition, $V'/V \sim 1/\sqrt{\Omega_Q}$, cannot be 
maintained.    A similar argument requires $1-(1-w_B)/(6+ 2 w_B)<\Gamma <1$ if 
$(1/2)(1+w_B)>w_Q> w_B$.
But, as we shall see, we also need 
that $\Gamma$ be nearly constant in order to have converging behavior.

If $\Gamma$ is nearly constant, then  the tracker equation,
Eq.~(\ref{Gamma}),  implies that  there is a solution in which $w_Q$
is nearly constant and $x$ and its time-derivatives are negligible. 
In this case, we have that the equation-of-state for the $Q$-field
is nearly constant:
\begin{equation}  \label{wrelate}
w_Q \approx \frac{ w_B - 2 (\Gamma -1)}{1+ 2 (\Gamma-1)}.
\end{equation}
If  $\Gamma >1$, the value of $w_Q$ must be less than $w_B$ to
satisfy the tracker equation,  which means that the $Q$-energy
red shifts more slowly than the background energy.
For inverse power-law potentials $V(Q) = M^{4+\alpha}/Q^{\alpha}$,
$\Gamma = 1 + \alpha^{-1}$, and the relation we have derived 
matches the relation in our first paper.\cite{zlatev1}
For $V(Q)= M^4 {\rm exp}(1/Q)$, $\Gamma= 1+ 2 Q$, which satisfies
our condition for $Q \ll 1$.
For $\Gamma<1 $, the same relation says
that the value of $w_Q $ must be greater than 
$w_B$; but we also require
that  $\Gamma >1-(1-w_B)/(6+2 w_B)$ in order to satisfy the
criterion, $w_Q < (1/2) (1+w_B)$, necessary to have converging
solutions (see Section III.F).

If $\Gamma$ varies significantly with $Q$, then one can find 
a wide and continuous distribution of
$Q$, $\Omega_Q$, and $w_Q$   which 
satisfies the tracker equation. 
Hence, there is no single solution to which solutions converge.

In sum,  we have succeeded in expressing our conditions for tracking
solutions in terms of $\Gamma$, 
which depends entirely on the functional form of $V$. 
A simple computation of $V''V/(V')^2$
determines if $V$ admits tracking solutions or not.
The case of interest for quintessence is $w_Q < w_B$, 
in which case the condition for a tracker solution 
is $\Gamma >1 $ and nearly constant.

\subsection{How is the tracker solution approached beginning
from different initial conditions?}

We shall call the tracker solution $\tilde{Q}(t)$ and the energy density
of the tracker solution as a function of time $\rho_{\tilde{Q}}(t)$.
This subsection  explains in rough detail
how solutions converge to the tracker solution for 
any initial $\rho_{Q}$ that lies between the initial background
energy density, $\rho_{Bi}$, and the current critical density,
$\rho_{c0}$. This  range stretches nearly 100 orders of magnitude.
In particular, we explain how 
the  convergence to the tracker solution is different if initially 
$\rho_B> \rho_Q > \rho_{\tilde{Q}}$ 
versus $\rho_{c0}<\rho_Q < \rho_{\tilde{Q}}$.
For simplicity, we confine ourselves to the case $w_Q <w_B$.
We complete the proof of convergence in the next subsection.

The equation-of-motion, Eq.~(\ref{teq}) can be rearranged as:
\begin{equation}  \label{eom2}
\frac{1}{6} \frac{d\, \ln{x}}{d\, \ln{a}} =-\frac{1}{3\sqrt{ \kappa  
(1+w_Q)}}\sqrt{\Omega_Q}\frac{V'}{V} -1 \equiv \Delta(t) -1;
\end{equation}
where we have restricted ourselves for simplicity to potentials 
with $V'<0$. Then,
since
$V' < 0$, the RHS of this equation  is a balance between a positive 
semi-definite 
term ($\Delta$) and a negative term. The tracker condition corresponds
to the two conditions:
 $1-w_Q^2$ significantly different from zero and
$w_Q$ nearly constant. The latter requires
near balance  on the RHS above ($\Delta \approx 1$) 
so that $d \, \ln{x}/d \, \ln{a}$ is
nearly zero.

\begin{figure}
 \epsfxsize=3.3 in \epsfbox{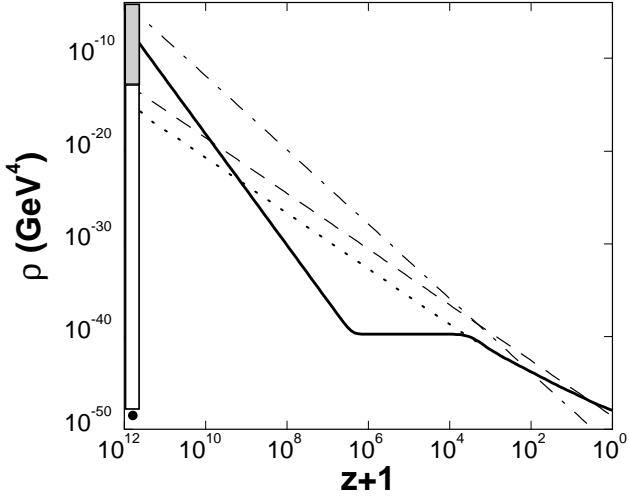}
 \caption{ 
Energy density versus red shift for the evolution of
 a tracker field.
 For computational convenience,
$z=10^{12}$ (rather than inflation)
 has been arbitrarily chosen as the initial
 time.
 The  white bar on left represents the range of initial
 $\rho_Q$
what leads to undershoot and  the grey bar represents overshoot,
  combining for a span of more than
   100 orders of magnitude if we extrapolate
   back to inflation.
 The solid black circle represents the unique initial condition
 required  if the missing energy is
   vacuum energy density.
	 The solid thick curve
	 represents an ``overshoot"   in which
	    $\rho_Q$ begins from a value greater than the tracker solution value,
	  decreases rapidly and freezes, and eventually joins the tracker
	      solution.
  }
  \end{figure}

First, consider the  ``overshoot" case in which 
$\rho_Q$ is initially much greater 
than the tracker value $\rho_{\tilde{Q}}$. For simplicity, let us assume
that $Q$ is released from rest.
The evolution goes through
four stages, illustrated in Figures~1-3.
The potential for this exampleis
   $V(Q)= M^4/Q^6$.
(For this and all subsequent figures,
 the choice of $z=10^{12}$ has been chosen as the initial
  time for convenience of computation and illustration; a realistic 
   figure would have initial $z$ corresponding to the inflationary scale.)
\begin{enumerate}
\item   $V'/V$ and $\Omega_Q$ are so big initially that  $\Delta \gg 1$.
 So, 
 \begin{equation}
 \frac16 \frac{d \, \ln{x}}{d\, \ln{a}} = \frac{2}{1-w_Q^2} 
 \frac{d\, \ln{w_Q}}{H dt} \gg 1
 \end{equation}
 and $w_Q$
 is driven towards its maximal 
 value,  $w_Q \rightarrow \, +1$.
This means
that $\dot{Q}$ becomes large  and $V$ decreases very rapidly as $Q$
runs downhill.  

\item As $Q$ runs downhill,
$V'/V$ and $\Omega_Q$ are decreasing.
Consequently, $\Delta$ begins to decrease and ultimately 
reaches a value of order unity,  one of the 
requirements for a tracker solution.
 However, 
 $w_Q$ has been driven towards $+1$;
up to this point, the RHS of Eq.~(\ref{eom2}) has been
positive, so there has been no opportunity for $w_Q$ to decrease.
 As a result, $1-w_Q^2$ is too small, or,
 more specifically, the kinetic energy
 is too large for $Q$ to join the tracker solution.
 Hence, $Q$ rolls
farther down the potential, overshooting the tracker solution.

\item Once the tracker solution is overshot, $\Delta$ becomes less
than unity and the RHS of Eq.~(\ref{eom2})
changes sign. $w_Q$
now decreases from $+1$ towards $-1$.  One might wonder what happens
when $w_Q$ crosses through
the tracker value; why doesn't $Q$ track at this point?
The answer is that there is now the problem that $\Delta$ is too
small. So, the RHS of Eq.~(\ref{eom2}) remains too negative and
$w_Q$ continues to decreases and heads towards $-1$.

\item Once $w_Q$ reaches close to $-1$, $Q$ is essentially frozen
at some value $Q_f$
and, consequently, $V$ and $V'/V$ are frozen. However, $\Delta$ is now
increasing since $\Omega_Q$ is increasing --  even though
$\rho_Q$ is nearly constant, $\rho_B$ is decreasing.  As
$\Delta$ increases to order unity, the sign of Eq.~(\ref{eom2})
changes once again; $w_Q$  increases from $-1$; the field runs
downhill; and,  the sign of Eq.~(\ref{eom2}) changes yet again. After
a few oscillations, the terms in Eq.~(\ref{eom2}) settle into
near balance and $Q$ is on track.

\end{enumerate}

  \begin{figure}
   \epsfxsize=3.3 in \epsfbox{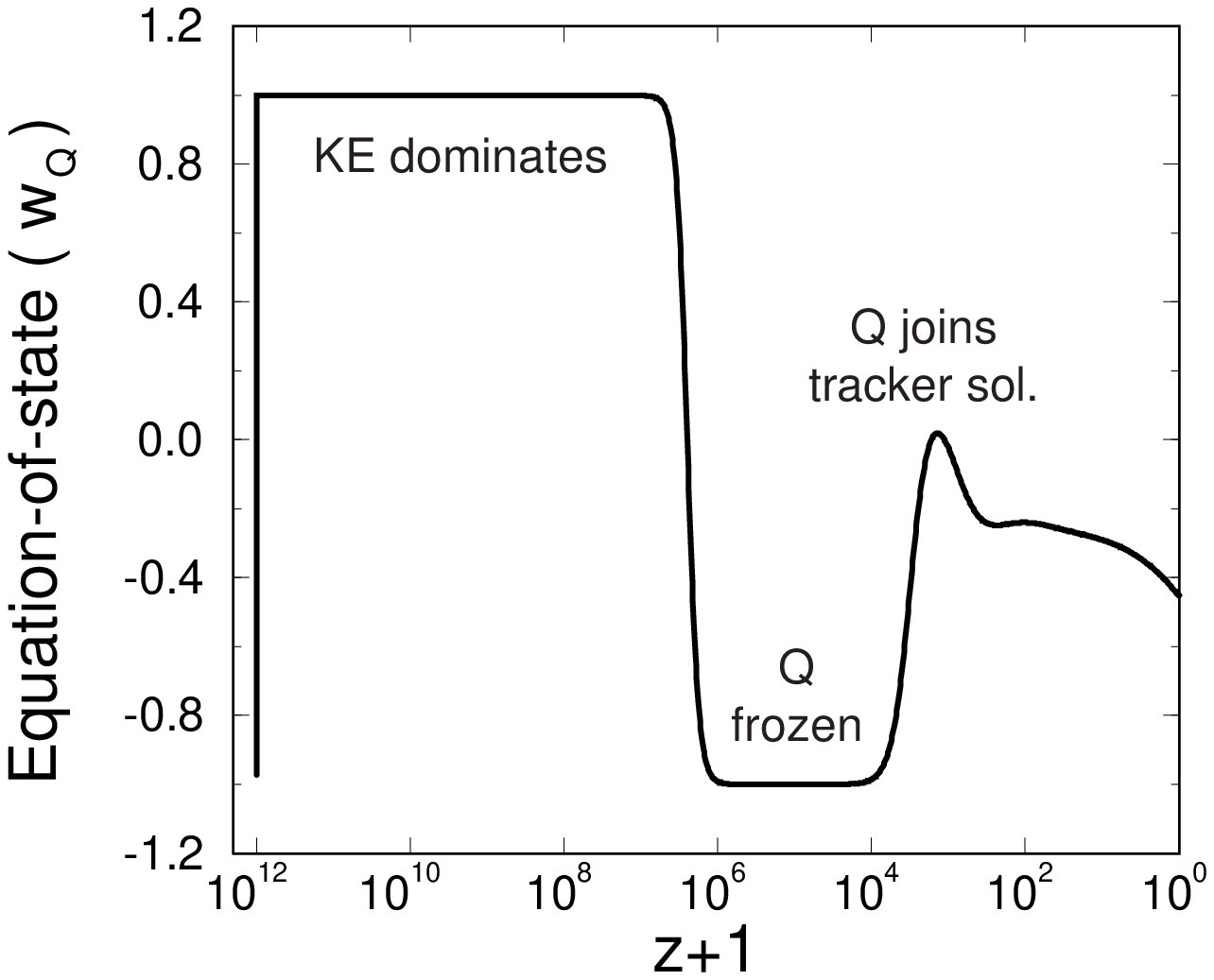}
   \caption{ A plot of $w_Q$ vs. red shift for the overshoot solution
   shown in Figure 1.  $w_Q$ rushes immediately towards $+1$ and
   $Q$ becomes kinetic energy dominated. The field freezes and
   $w_Q$ rushes towards $-1$. Finally, when $Q$ rejoins the tracker
   solution, $w_Q$ increases, briefly oscillates and settles into
   the tracker value.}
   \end{figure}

   \begin{figure}
    \epsfxsize=3.3 in \epsfbox{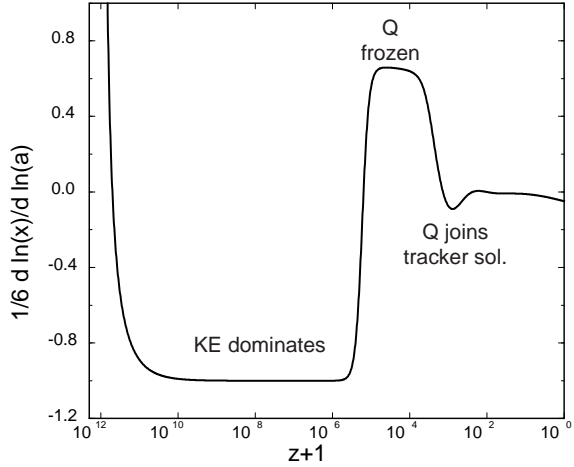}
    \caption{ A plot of $\dot{x}/6 =(1/6)\,  d \, \ln{x}/d\, \ln{a}$
    for the overshoot
    solution shown in Figures 1 and 2.  At late times when
    $Q$ settles into the tracker solution, $\dot{x}$ is small
    and $w_Q$ is nearly constant. During the
    overshoot phase, $\dot{x}$ undergoes large positive and
    negative changes, as described in the text.}
    \end{figure}

 Next, consider the ``undershoot" case in which $\rho_Q$ is initially much
 less than the tracker value $\rho_{\tilde{Q}}$ and $Q$ is
 released from rest. This corresponds to
   $Q \gg \tilde{Q}$ initially.  
By assumption, $V$ and $|V'/V|$ are much smaller than the tracker value.  
Consequently, $1/\sqrt{\Omega_Q}$ is larger than the tracker value.
The only way to satisfy the  equation-of-motion, Eq.~(\ref{teq}) is
for $w_Q$ to approach $-1$ so that the coefficient of $1/\sqrt{\Omega_Q}$
is nearly zero. This condition corresponds to a very small 
kinetic energy density or $Q$ nearly constant. Hence, the field 
remains nearly ``frozen", and $V$  and $V'/V$ are
nearly constant as the universe
evolves.  The situation is identically to beginning with Step 4 above, and
the scenario proceeds just as described there.

In sum, the field either drops precipitously past the tracker
value and is frozen (overshoot), or it begins with 
a value less than 
the tracker solution (undershoot) and is frozen.  
In either case, it 
proceeds from the frozen state to joining the tracker solution.
In the case of undershoot, the frozen value $Q_f$ is simply the 
initial value of $Q$.
For the overshoot case, $Q$ begins by going through a
kinetic energy dominated period in which $\dot{Q} \propto a^{-3}$.
If the initial $\rho_{Qi} < \rho_{Bi}$, where $\rho_B$ is the background
radiation density, then $a \propto t^{1/2}$ and 
the frozen value of $Q$ is 
\begin{equation} \label{freeze}
Q_f \approx Q_i + \sqrt{\frac{3}{4 \pi} 
\Omega_{Q_i}}
\end{equation}  
where the subscript refers to the initial values of $Q$ and $\Omega_Q$.
(If  initially  $\rho_{Qi} > \rho_{Bi}$, then $a \propto t^{1/3}$, 
and 
\begin{equation}
Q_f \approx Q_i + \sqrt{\frac{3 }{4 \pi}} (1+\frac{1}{2} 
 \ln{\frac{\rho_{Qi}}{\rho_B}});
\end{equation}
in this case, $Q_f$ is so large that $Q$ remains frozen 
up to the present time.)
For the overshoot case, $Q_i$ is typically very small compared to 
unity and $\Omega_{Qi} = {\cal O}(1)$.
Consequently, the frozen value $Q_f$ depends  on $\Omega_{Qi}$
only.  

Initial conditions in which $\dot{Q}_i$ is non-zero do not change the
story significantly.  If $\dot{Q}_i$ is very large, then the initial 
behavior is kinetic energy dominated, and the evolution proceeds 
similar to the overshoot case.  
Initial fluctuations in $Q$ also
do not change the story since they are exponentially suppressed once
the potential becomes non-negligible and the field is driven towards
the tracker solution (see Appendix).

The possibility of overshoot and undershoot allows a new possibility for
the case of exponential potentials 
  recently discussed by Ferreira and
  Joyce.\cite{Ferreira}
  The exponential potential is a special example of a tracker
  solution 
  in which $\Omega_Q$ is constant during the matter dominated epoch.
The  practical problem with this model, as noted by Ferreira and Joyce,
  is $\Omega_Q$ is constrained to be small ($\Omega_Q \le 0.15$).
   At the beginning of matter domination, $\Omega_Q$ must be small
   in order that large-scale structure be formed; but then it cannot
   change thereafter. Hence, it remains a small, subdominant
   component. This argument presumes, however, that $Q$ is already
   on track at the beginning of matter domination. It is possible
   to tune initial conditions so that $Q$ overshoots the tracker
   solution  initially and does not join the tracker solution until
   just very recently (red shift $z=1$).  Then, the constraint on $\Omega_Q$
   is lifted.

\subsection{What are the constraints on the initial value of $Q$ and
$\rho_Q$?}

Suppose that the tracker solution corresponds to $Q=Q_0$ today  and
the $Q$ has converged to a tracker solution.
Then, whether $\rho_Q$ is initially smaller than the tracker value
and frozen at some $Q= Q_f$ equal to its initial value,
or $\rho_Q$ is initially larger than
the tracker value and falls to $Q_f \approx \sqrt{3 \Omega_{Qi}/4 \pi}$,
it is necessary that $Q_f$  be less than $Q_0$ in order that the field 
be tracking today.  This is not a very strong constraint.  
Since $Q_0 = {\cal O}(1)$ for most tracking 
potentials, this only requires that 
 $\rho_{c0} < \rho_{Qi} <  \rho_{Bi}$ initially.  Hence,
initial conditions in which $Q$ dominates the radiation and matter
density
are disallowed because $Q$ falls so fast and drops to such a low point on the potential
that it has not yet begun tracking today.
  However, initial
conditions in which  there is rough equipartition between $\rho_Q$ and
the background energy density are allowed, as well as initial
values of $\rho_Q$ ranging as low as 100 orders of magnitude 
smaller, comparable to the current matter density.
The allowed range is impressive and spans the most physically likely
possibilities.

\subsection{Is the tracker solution stable?}

What has been shown so far is that, whether the initial conditions
correspond to undershoot or overshoot, $Q$  soon reaches some frozen
value $Q_f$ which depends on the initial $Q_i$.
Then, after some evolution, $\Omega_Q$ increases to the point
where $|V'/V| \sim 1/\sqrt{\Omega_Q}$ and, according to the 
equation-of-motion, $w_Q$ moves away from $-1$ and the field 
begins to roll.  
What remains to be shown is that 
solutions with $w_Q$ not equal to the tracker solution value
converge to the tracker solution. Or, equivalently,
we need to show that the tracker solution is stable.

Now, consider a solution in which $w_Q$ differs from the tracker 
solution value $w_0$  by
an amount $\delta$. Then, the master equation can be expanded to lowest
order in $\delta$ and its derivatives to  obtain after some algebra:
\begin{equation}
\ddot{\delta}
+3[\frac12 (w_B+1)-w_0]\dot{\delta}
+\frac{9}{2}(1+w_B)(1-w_0)\delta = 0
\end{equation}
where  the dot means  $d/d\, \ln{a}$ 
as in the tracker equation.
The solution of this equation is
\begin{equation}
\delta \propto a^{\gamma}
\end{equation}
where
\begin{eqnarray}  \label{gam} 
	&& \gamma = -\frac{3}{2}[\frac12 (w_B+1)-w_0] \nonumber \\ 
	&& \pm \frac{i}{2}
	\sqrt{18(1+w_B)(1-w_0)-9[\frac12 (w_B+1)-w_0]^2   }.
\end{eqnarray}
The real part of
the exponent  $\gamma$ is negative for $w_0$ between  $-1$ and
$w_0 = \frac{1}{2}(1+w_B)$, which includes our entire range of interest.
So, without imposing any further conditions,
this means that $\delta$ decays exponentially and the solution
approaches the tracker solution.
As $\delta$ decays, it also oscillates with a frequency described
by the second term. See Figure 4.

 \begin{figure}
  \epsfxsize=3.3 in \epsfbox{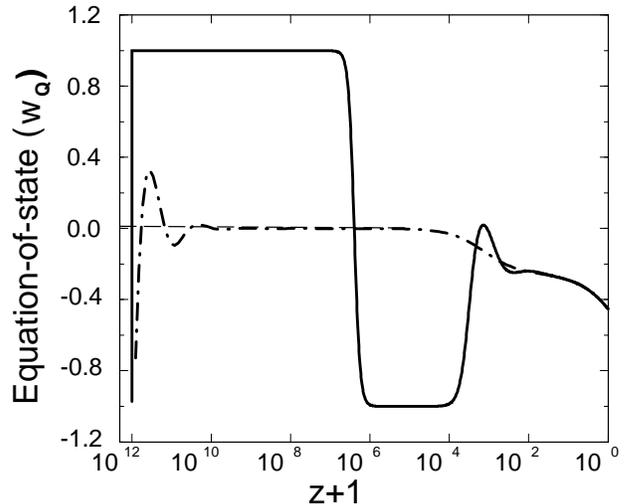}
  \caption{  The convergence of different initial conditions
  to the tracker solution. As derived in the text, $w_Q$
  decays exponentially fast to the tracker solution combined
  with small oscillations. All the curves are for $V(Q)= M^4/Q^6$.
  The solid curve is the overshoot case from Figure 1. The thin
  dashed curve  with $w \approx 0$ is
   the tracker solution which is overlaid for most $z$ by 
   the dash-dotted curve, which represents    a slightly
  undershooting solution.}
  \end{figure}

In deriving Eq.~(\ref{gam}),
we have assumed that $\Gamma$ is strictly constant, independent
of $Q$, which is exactly true for pure inverse power-law 
($V\sim 1/Q^{\alpha}$) or exponential ($V\sim {\rm exp}(\beta Q)$)
potentials.
The same result holds if 
$|d (\Gamma-1)/H d t| / |\Gamma -1| \ll 1$
({\it i.e.},  $\Gamma-1$ varies with $Q$ but only by a modest
amount)   over
the plausible range of initial conditions ranging from
 $V(Q) \approx \rho_{Bi}$ to $V(Q) \approx \rho_{eq}$ (where
 $\rho_{Bi}$ is the initial background energy density after inflation,
 say, and 
$\rho_{eq}$ is the energy density at matter-radiation equality).
The condition 
is equivalent to $| \Gamma'/[\Gamma \, (V'/V)] |\ll 1$.
In this limit, 
$\Gamma$ and 
the tracker value of $w_Q$ change adiabatically as $Q$ rolls downhill,
satisfying the tracker equation 
with $\dot{x}$  being negligibly small, as discussed 
in the Appendix.
The constraints on $w_Q$ are the  same as above. However,
some important differences from the constant $\Gamma$ case
are pointed out in the last section.

Throughout most of our discussion in this paper, we have 
considered the case $w_Q< w_B$. However,
our convergence 
criterion, $w_0 < \frac{1}{2}(1+w_B)$, includes $w_Q > w_B$
or, equivalently,  $1- (1-w_B)/(6+2 w_B)< \Gamma <1$, as also
found by Liddle and Scherrer.\cite{LidSher}
An example is $V \sim Q^{\alpha}$ with 
$\alpha \ge (6+2 w_B)/(1-w_B)$; $\alpha \ge 6$ for $w_B=0$ and
$\alpha \ge 10$ for $w_B=1/3$.
Let us suppose we reached the present $\Omega_Q$ after tracking
down this potential.
Because $w_Q > w_B$ in these potentials, 
it must be that $\Omega_Q$ exceeds $\Omega_B$ extrapolating backwards
in time.  Consequently,
there is no period of matter-domination or structure 
formation, and  these models have no practical interest.
However, see the discussion of hybrid models below for a variation
on these models that may be viable.

\subsection{Borderline models and hybrid models}

For completeness, we consider two special classes of potentials, 
borderline trackers in which $\Gamma =1$ and hybrid models in which 
$\Gamma >1$  at first and then $\Gamma <1$.

The borderline case corresponds to $V \propto {\rm exp}(\beta Q)$, 
which has been studied by several authors.\cite{Ratra,Wett,Ferreira,Cope}
For this case, the tracker equation for $\dot{x}=\ddot{x}=0$ 
demands that $w_Q= w_B$ and,
therefore,
$\Omega_Q$ is constant. Hence, for this case, the tracker solution 
corresponds to maintaining a constant ratio of quintessence to background
energy density.  The only deviation occurs during the transition from 
radiation- to matter-domination when $\dot{x}$ becomes non-negligible,
but this is a small effect.  

Because $\Omega_Q$ is constant throughout the matter-dominated epoch, 
these models have limited practical utility.
$\Omega_Q$ must be small ($\le 15\%$) at the onset of matter-domination
in order not to disrupt structure formation.  (Quintessence suppresses
the growth rate.)  But, then, since $\Omega_Q$ is constant, $\Omega_Q$
remains small forever.  Consequently, the models require $\Omega_m > 85\%$,
inconsistent with a number of determinations of mass\cite{Wang98b}, and
the universe never enters a period of accelerated expansion, inconsistent
with recent measurements of the luminosity-red shift relation for
Type IA supernovae.\cite{supernova}
(The overshoot scenario may lift the $\Omega_m > 85\%$ constraint
 but, as discussed in Section D, introduces fine tuning which 
 defeats the whole purpose
 of the scenario.)

Hybrid models have the property that solutions converge to a tracker solution
at the early phase of evolution but cease to converge after a certain
point due to a change in the shape of the potential as $Q$ rolls downhill.
One can imagine a sufficiently long convergence
regime that all or most plausible 
initial conditions have collapsed to a common tracker solution
before the second regime begins.  Effectively, this has the desired 
feature that a wide range of initial conditions lead to the same final
condition.  The models may be   somewhat artificial
in that the current cosmology
is very sensitive to where the transition 
occurs. For example, consider the case where $w_Q < w_B$ but $\Gamma$
undergoes a transition from $\Gamma >1$ (converging) to $\Gamma <1 $.
Recall that $\Gamma <1$  corresponds
to $|V'/V|$ increasing as $V$ decreases. We have shown that, extrapolating
backwards only a small interval in time, the field $Q$ 
must have been frozen at a value not
so different from the current value (assuming the field is rolling today and 
$w_Q<0$). In this kind of hybrid model, the transition
to $\Gamma <1$ must be set so  that the transition 
occurs so that $Q$ is near the 
frozen value, which requires delicate tuning of parameters in the 
potential.

A different example is where $1- (1-w_B)/(6+2w_B)<\Gamma <1$ and $w_Q >w_B$
during the early stages
of the universe. We have argued that these conditions produce
converging behavior but, if the conditions continue to the present,
there is no period of matter-domination or structure 
formation (see Section III.F).
However, one can imagine hybrid models in which these 
conditions $[1- (1-w_B)/(6+2w_B)]<\Gamma <1$
and $w_Q > w_B$ are satisfied
for some period early in the history of the universe, 
providing a finite period of converging behavior.  Later, as $Q$
moves down the potential, $V$ changes form so that $w_Q < w_B$.
Viable cosmological models 
of this type can be constructed in which $\Omega_Q$ does
not dominate the universe during the matter-dominated epoch until
near the present time.

\subsection{Some additional practical considerations}

We have discovered a wide class of potentials that exhibit tracking behavior.
This guarantees that a wide range of initial conditions converge
to a common tracking solution, but  the convergence may take longer than 
the age of the universe in some cases.   In particular, if
one assume equipartition after inflation, say,  and the initial
$V$ is too far above the tracker solution, then $Q$ falls precipitously,
overshoots the tracker solution, and freezes at some $Q=Q_f$.
For some potentials satisfying the condition
$\Gamma \approx \, {\rm constant} \, >1$,
$Q_f$  may be so large that
 the field does not begin to roll and track
by the present epoch.  If $Q$ just started to roll by the present
epoch, then it would behave exactly as a cosmological constant until
now, and so the model is trivially equivalent to a $\Lambda$ model.
As a practical consideration, we  demand that a field starting from
equipartition initial conditions should start
rolling by matter-domination, say, so that the model is non-trivial. 
This imposes a mild added constraint on potentials, $V(Q)$.
For this purpose, rough estimates suffice.

Equipartition at the end of inflation, when there are hundreds or
perhaps thousands of degrees of freedom in the cosmological
fluid, means that the $Q$-field
has $\Omega_i \approx 10^{-3} $.  Beginning from equipartition, 
$Q$ falls to some value $Q_f$ where it freezes.  According to 
the  equation-of-motion, $Q$ remains frozen
until $\sqrt{\Omega_Q} \propto \sqrt{V}/H$ increases to where
$V'/V \sim 1/\sqrt{\Omega_Q}$ or, equivalently,
\begin{equation}
        \frac{V'^2}{V}(Q_f) > H^2(z_{eq})=  \frac{H_0^2}{a^3_{eq}}
        \sim \frac{1}{\Omega_0} \frac{V_0}{a^3_{eq}}.
\end{equation}

For  $V(Q) \propto 1/Q^{\alpha}$,  this imposes the constraint
\begin{equation}
\frac{\alpha^2}{Q_{f}^{\alpha +2}}
\geq \frac{1}{\Omega_0 \, Q_{0}^{\alpha } \, a^{3}_{eq}},
\end{equation}
where $Q_0$ is the present value of $Q$.
Since today
\begin{equation}
        \frac{V_0'^2}{V_0} \sim H_0^2 \sim \frac{8 \pi}{3}
        \frac{V_0}{\Omega_0},
\end{equation}
we obtain
$Q_0^2 \sim \frac{3}{8 \pi} \Omega_Q^0 \alpha^2$.
From Eq.~(\ref{freeze}), we also know that
\begin{equation}
        \frac{4 \pi}{3} Q_{f}^2 \sim 
	\Omega_i \sim 10^{-3} .
\end{equation}
Combining the above relations one gets the restriction on $\alpha$
\begin{equation}
	\alpha^2 > \frac{10^{12}}{\Omega_0} \frac{(\frac{3}{4 \pi}\Omega_i)
	^{(1+\alpha/2)}}{(\alpha^2 \Omega_0 \frac{3}{8\pi})^{\alpha/2}}
\end{equation}
where we have taken $a_{eq} \sim 10^{-4}$. This approximate relation leads
to $\alpha \ge  5$. Figure~5 confirms this result showing that
the $\alpha =1$ model  starts rolling much later than equality. 
So, if one
restricts  to  pure inverse power-law ($1/Q^\alpha$)
potentials, our constraint that $Q$ begin from equipartition and roll before
matter-radiation equality constrains us to
$\alpha \ge  5$.

\begin{figure}
 \epsfxsize=3.3 in \epsfbox{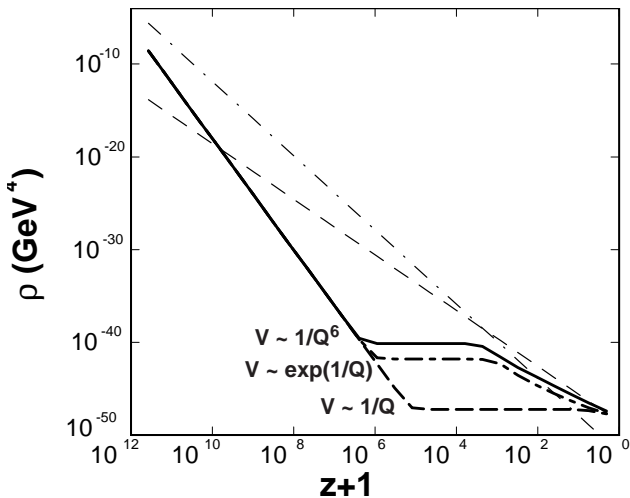}
  \caption{A comparison of the overshoot for
    three different models beginning from
     $\Omega_i = 10^{-3}$ (equipartition).
      The thick solid line is for $V(Q)= M^4/Q^6$, 
       the thick dash-dotted line is for
	$V=M^4[exp(1/Q)-1]$,  and the
      thick long dashed line is for
       $V(Q)= M^4/Q$.
       In all three examples, $Q$ falls rapidly downhill and freezes.
	In the first and second
	 examples, $Q$ begins to roll again and joins the
	  tracker solution  before matter-radiation
	   equality; the third example, which violates the condition
	    derived in the text, does not begin to roll again by the
	     present epoch.
	      }
	       \end{figure}

\section{The $\Omega_Q$-$w_Q$ relation}

An extremely important aspect of tracker solutions is the 
$\Omega_Q$-$w_Q$ relation, or, equivalently, the 
$\Omega_m$-$w_Q$ relation which it forces.
For any given $V=M^4 \tilde{v}(Q/M)$ (where $\tilde{v}$ is
a dimensionless function of $Q/M$), $Q$ and $\dot{Q}$ are 
totally determined independent of initial conditions by the 
tracker solution.
The only degree of freedom is the $M$-parameter in the potential.
$M$ can be fixed by imposing the constraint that the universe is 
flat and $\Omega_m=1-\Omega_Q$ is determined by measurement.
There is, then, no freedom left to independently vary $w_Q$.
This is the explanation of the $\Omega_Q$-$w_Q$ relation,
a new prediction that arises from tracker fields.
The $\Omega_Q$-$w_Q$ relation is not unique because there
remains the freedom to change the functional form of $V(Q)$.
Even so, $w_Q$ is sufficiently constrained as to be cosmologically
interesting.

The general trend is that $w_Q \rightarrow -1$ as $\Omega_Q \rightarrow
1$. The fact that $\Omega_m \ge 0.2$ observationally means
that $\Omega_Q \le 0.8$ and $w_Q$ cannot be very close to $-1$.
How small $w_Q$ can be is model-dependent.  We are most interested
in the smallest values of $w_Q$ possible since the difference
from $-1$ determines how difficult it is to distinguish the tracking
field candidate for missing energy from cosmological constant.

In Figure~6, we show the $\Omega_Q$-$w_Q$ relation for 
a series of pure, inverse power-law potentials, $V(Q) \propto
1/Q^{\alpha}$.  The general trend is that $w_Q$ increases
as $\alpha$ decreases.  The constraint given at the end of 
the previous section is that $\alpha \ge 5$ (in order that 
$Q$ be rolling by matter-radiation equality beginning from
equipartition initial conditions).  For $\Omega_Q=0.8$,
the smallest value of
$w_Q$ is -0.52, which occurs for $\alpha=5$.
This is a large difference from $-1$ obtained for a cosmological
constant.

However, a lower value of $w_Q$ can be easily achieved for a
more generic potential with a mixture of inverse power-laws,
{\it e.g.}, $V(Q) = M^4 \, {\rm exp}(1/Q)$.   For these 
models, $Q$ is small initially.
If the potential is expanded in inverse powers of $Q$, $1/Q^{\alpha}$, then it is dominated in the early stages by the high-$\alpha$
terms.  Hence, the effective value of $\alpha$ is much greater
than 5 before matter-radiation equality, and we easily satisfy
the constraint that the field be rolling before matter-radiation
equality beginning from equipartition initial conditions.
On the other hand, the value of $Q$ at the present epoch is large,
and the potential is dominated by the $\alpha \approx 1 $ terms in 
its expansion.  Consequently, $w_Q$ can be even lower
today than in the pure power-law case.  For $\Omega_Q=0.8$, we
obtain $w_Q=-0.72$ for the exponential potential, which is
in better accord with recent constraints on $w_Q$ from
supernovae.\cite{supernova}

It is difficult to go below this limit without artificially
tuning potentials unless we relax our constraints.
For example,  consider the highly contrived potential 
$V(Q)= \frac{A}{Q^{10^{-5}}} + \frac{B}{Q^{10}}$, in which we 
have intentionally chosen exponents differing by six orders of
magnitude in order to obtain a small $w_Q$ today. The
second term in the potential dominates before equality and insures
that the field is rolling by that point in time.
The first term dominates at late times and
makes the equation of state $w_0$ very low (because this term is very flat).
For this example, we find  
$w_Q = -0.98$ for $\Omega_Q=0.8$. However, we had to choose 
a pair of terms with exponents differing by six orders of 
magnitude.
The lesson of this exercise  (and related tests) is that the
exponential potential is a reliable estimate for the minimal $w_Q$
possible for generic, untuned potentials.

\begin{figure}
 \epsfxsize=3.3 in \epsfbox{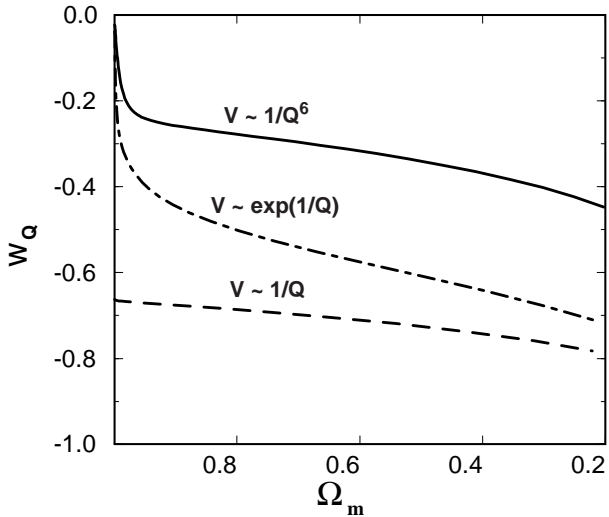}
  \caption{The  $\Omega_Q$-$w_Q$ relation  for various potentials 
  assuming a flat universe $\Omega_m=1-\Omega_Q$,
  where $w_Q$ represents the present value of $w_Q$. The
  potentials and notation are the same as in Figure 5.
   }
    \end{figure}

In Figure 6, we illustrate the  $\Omega_Q$-$w_Q$ relation
for several potentials.   Only the $V\sim 1/Q^6$ and 
$V\sim {\rm exp}(1/Q)$ potentials satisfy the condition 
that $Q$ is rolling by matter-radiation equality beginning
from equipartition initial conditions.
In Figure 7, we illustrate the effective $w_Q^{eff}$
that would be measured using supernovae or cosmic microwave
background measurements,  
using the $V\sim {\rm exp}(1/Q)$ potential as defining the boundary 
of minimal values of $w_Q$ possible for the tracker field case.
This boundary assumes that we satisfy the strict condition that
the field  be rolling by 
matter-radiation equality beginning from equipartition initial 
conditions.  If we relax
this condition and allow a somewhat narrower range of initial
conditions,  then general  potentials of
the form
       $V \sim \Sigma c_k/Q^k$ 
      (such as $V \sim 1/Q$) are allowed
and $w_Q$ can be somewhat smaller (see Figure 6). Hence, in 
Figure~7,  
the boundary can relax somewhat downward (dashed line)
but it is difficult
to obtain $w_Q < -0.8$ or $w_Q^{eff}<-0.75$.
Because $w_Q$ is evolving at recent 
times, the value obtained from measurements at moderate to 
deep red shift will differ from the current value shown in 
Figure 6.  For tracker potentials, the effect of integrating
back in time over varying $w_Q$ turns out to be well-mimicked
by  a model with constant $w_Q = w_Q^{eff}$   
that has the same conformal distance to last 
scattering surface.
 For the case $\Omega_m=0.2$, for example, Figure 6
shows that $w_Q=-0.72$ today, but Figure 7 shows that the 
measured $w_Q$ would be $w_{eff} =-0.63$.

The result is exciting because the $\Omega_Q$-$w_Q$ relation
and the constraint on $\Omega_m \ge 0.2$ today  creates 
a sufficiently large gap between $w_Q^{eff}$ and $-1$ that
the tracker candidate for missing energy should be distinguishable
from the cosmological constant in near-future cosmic
microwave background and supernovae measurements.

\begin{figure}
 \epsfxsize=3.3 in \epsfbox{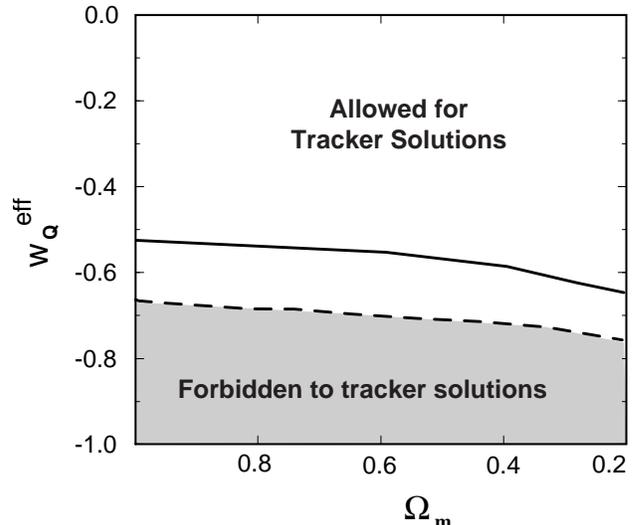}
  \caption{A  plot of $w_Q^{eff}$ versus
  $\Omega_m= 1-\Omega_Q$, 
    showing the minimum $w_Q^{eff}$ possible for 
   tracker solutions. The solid line is the lower boundary
   assuming  the constraint that the $Q$-field 
   begins with equipartition initial conditions and begins rolling
   before matter-radiation equality. 
   The dashed line is the  lower boundary if this condition 
   is relaxed to allow general $V \sim \Sigma c_k/Q^k$.
     As explained in the text, $w_Q^{eff}$ is the value
    that would be measured in supernovae and microwave background
     experiments
     which effectively integrate over a varying $w_Q$.
	}
	 \end{figure}

\section{Why is the universe accelerating today?}

We have proven in this paper that tracker potentials resolve the
coincidence problem for quintessence.   For a very wide range of
initial conditions, cosmic evolution converges to a common track. The 
tracker models are similar to inflation in that they funnel a 
diverse range of initial conditions into a common final state.
The models have only one important free parameter ($M$) which is 
fixed by the measured $\Omega_Q= 1- \Omega_m$.  

Some of the mathematical  properties of tracking solutions
have been noted before for ${\rm exp}(\beta Q)$ 
and pure, inverse power-law potentials.\cite{Ratra,Wett,Ferreira,Cope}
However, the extraordinary insensitivity to initial conditions and
the potential application to the coincidence problem was not 
explored.
The present work is important
because it shows that the properties are shared by a much wider 
class of more generic potentials.
``Generic potentials"  include, for example,
all $V$'s  which can be
expanded as a finite or infinite  sum of terms with  inverse
powers of $Q$, which is much more general
than the special cases of a single inverse-power
or a pure exponential. We use $V= M^4 \, {\rm exp}(1/Q)$ as an 
example of this more generic class, although our conclusions would
remain the same for more general $V= \Sigma c_k/Q^k$.

Extending the tracker behavior to generic potentials may 
be important because, as we shall argue below,
they have properties not shared by 
 the special cases 
($V \sim 1/Q^{\alpha}$ and $V \sim {\rm exp}(\beta Q)$)
that make plausible why
$\Omega_Q$ only begins to 
to dominate   and initiate a period of 
accelerated expansion late in the history of the universe.

To understand this conclusion, we need to change our approach.
Up to this point in the paper, we have imagined fixing $M$ to 
guarantee that $\Omega_Q = 1- \Omega_m$ has the measured value today.
This amounts to considering one tracker solution for each $V(Q)$.
Now we want to consider the entire family of tracker solutions 
for each given $V(Q)$ and consider 
whether $\Omega_Q$
is more likely to dominate late in the  universe for one $V(Q)$
or another.

\begin{figure}
 \epsfxsize=3.3 in \epsfbox{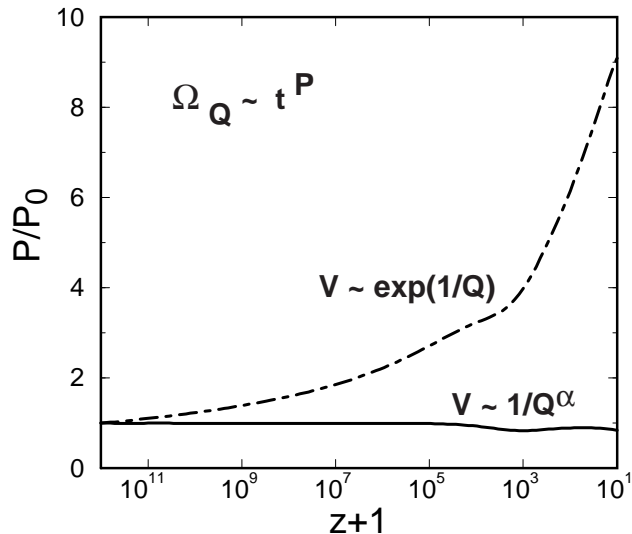}
  \caption{
A plot of $P/P_0$ versus $t$, where $\Omega_Q \propto t^P$ and 
$P_0$ is the initial value of $P$.
The plot compares  pure inverse power-law ($V\sim 1/Q^{\alpha}$)
potentials for which $P$ is constant with a generic potential
 ({\it e.g.},  $V \sim {\rm exp}(1/Q)$)
for which  $P$
increases with time.
	}
	 \end{figure}

In general, $\Omega_{Q}$ is proportional to
$a^{3(w_B-w_Q)} \propto t^{2(w_B-w_Q)/(1+w_B)}$, where we have shown 
in  Eq.~(\ref{wrelate}) that
\begin{equation}
w_B-w_Q = \frac{2(\Gamma-1) (w_B+1)}{1+2(\Gamma-1)}.
\end{equation}
Hence, we find $\Omega_Q \propto t^P$ where 
\begin{equation}
P= \frac{4 (\Gamma-1)}{1+2(\Gamma-1)}.
\end{equation}

A satisfying explanation as to why $\Omega_Q$ 
dominates at late times, rather than early, would be if $\Omega_Q$
automatically changes behavior at late times.
This is not the case
for the two special cases ($V\sim 1/Q^{\alpha}$ and
$V \sim {\rm exp}(\beta Q)$). As the universe evolves, 
$\Gamma-1$ and, hence, $P$  are  constant;
consequently,  $\Omega_Q$
grows as the same function of time throughout the radiation- and
matter-dominated epochs.
However, these are the  exception, rather than the rule.

For more general potentials, $P$ increases as the universe
ages. Consider first a potential which is the sum of two inverse power-law
terms with exponents $\alpha_1 < \alpha_2$.  The term with the 
larger power is dominant at early times when $Q$ is small, but the term with the smaller
power dominates at late times as $Q$ rolls downhill and obtains a larger value.
 Hence, the effective value of $\alpha$
decreases and $\Gamma-1 \propto 1/\alpha $ increases; the result is
that $P$ increases at late times.   For more general potentials,
such as $V \sim {\rm exp}(1/Q)$, the effective value of $\alpha$ decreases
continuously and $P$ increases with time. 
Figure~8 illustrates the comparison in the growth of $P$.

How does this explain why $\Omega_Q$ dominates late in the 
universe?  Because an increasing $P$ means that $\Omega_Q$ 
grows more rapidly as the universe ages.
Figure~9 compares a tracker solution for a pure inverse power-law
potential ($V \sim 1/Q^6$) model
with  a  tracker solution for $V \sim {\rm exp}(1/Q)$, where the
two solutions have been chosen to begin at the same value of 
$\Omega_Q$.
 (The start time has been chosen
arbitrarily at $z = 10^{17}$ 
for the purposes of this illustration.)
Following each curve to the right, there is a dramatic
(10 orders of magnitude) difference between 
the time when  the first solution (solid line) meets the background density
versus the second solution (dot-dashed line).
That is, beginning from the same
$\Omega_Q$, the first tracker solution dominates well before matter-radiation
equality and the second (generic) example dominates well after
matter-domination.
The difference is less dramatic as $\alpha$ increases for the pure 
inverse power-law model and becomes negligible for 
$\alpha > 15$.  Of course, the model appears more contrived.
But, more importantly,
as $\alpha$ increases, the value of $w_Q$ today
(given $\Omega_m \ge 0.2$) approaches zero and the universe does
not enter a period of acceleration by the present epoch.  
Hence, a significant conclusion is that the pure exponential
and inverse power-law
models are atypical; the generic potential has properties that 
make it much more plausible that $\Omega_Q$ dominates
late in the history of the universe and induces a recent period of 
accelerated expansion.

\begin{figure}
 \epsfxsize=3.3 in \epsfbox{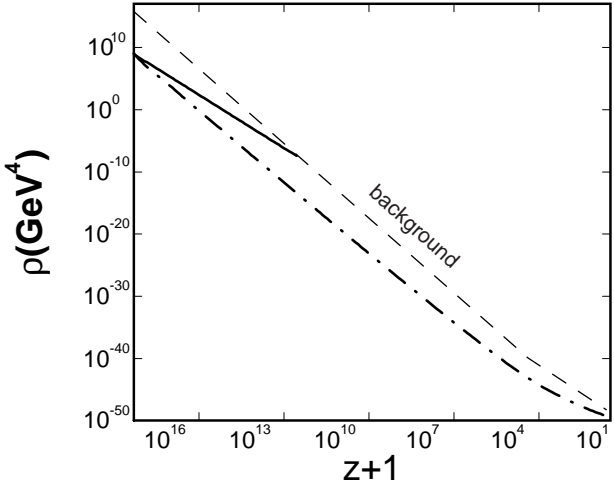}
   \caption{ A plot  comparing two tracker solutions for the
   case of a $V \sim 1/Q^6$ potential (solid line) and 
   a $V \sim {\rm exp}(1/Q)$ potential (dot-dash). The dashed
  line is the background density. The two tracker solutions
  were chosen to have the same energy density initially. The
  tracker solution for the generic example ($V \sim {\rm exp}(1/Q)$)
  reaches the background density much later than for the pure
  inverse-power law potential.  Hence, 
$\Omega_Q$ is more likely to dominate late in the history of the 
universe in the generic case.
	   }
		    \end{figure}

In sum,   the general tracker behavior shown in
this paper 
goes a long way towards resolving two key issues: 
the coincidence (or initial conditions) problem  and why $\rho_Q$ 
is dominating today 
rather
than at some early epoch. And, it leads to a new prediction -- 
a relation between $\Omega_m=1-\Omega_Q$ and $w_Q$ today that makes
tracker fields distinguishable from a cosmological constant.

We wish to thank Robert Caldwell  and Marc Kamionkowski
 for many insightful comments and suggestions.
This research was supported by the US Department of Energy grants
DE-FG02-92-ER40699 (Columbia)
 DE-FG02-95ER40893 (Penn) and
 DE-FG02-91ER40671 (Princeton).

\section{Appendix}

In this Appendix, we discuss the convergence to the tracker
solution when $\Gamma \equiv V''V/(V')^2$ varies with $Q$. 
We wish to show that convergence occurs if the variation of $\Gamma$ 
over the plausible range of initial conditions (varying of 100 
orders of magnitude in energy density) is nearly constant.
An example is $V \sim {\rm exp}(1/Q)$ for which $\Gamma = 
1+2 Q$ and $Q \ll 1$ for the plausible range of initial conditions.

The condition that $\Gamma$ is nearly constant means precisely
that $d (\Gamma-1)/Hdt \ll (\Gamma-1)$ or, equivalently,
$| \Gamma'/[\Gamma \, (V'/V)] |\ll 1$.
In this case, $\Gamma$ is nearly constant over a Hubble time.
Hence, we can consider an adiabatic approximation 
for the tracker solution in which 
$Q_0$ and $w_0$ satisfy the tracker equation Eq.~(\ref{Gamma})
with $\dot{x}$ and $\ddot{x}$
negligibly small. Suppose $w_Q$ and $Q$ are both perturbed from this 
tracker solution by amounts $\delta w$ and $\delta Q(t)$.
From the definition of $w_Q$, we have that
\begin{equation} \label{eq1}
        \delta w = (1-w_0)\frac{\dot Q_0}{\rho_0} \delta\dot{Q}
        - (1+w_0)\frac{V_0'}{\rho_0}\delta Q.
\end{equation}
From the equation-of-motion Eq.~(\ref{teq})  we know that
\begin{equation}
        \frac{V_0'}{\rho_0}  
	\sim \frac{V'_0}{V_0} \propto 
	\frac{1}{\sqrt{\Omega_Q}} \propto a^{-3(w_Q-w_B)}.
\end{equation}
Consequenty, Eq.~(\ref{eq1}) implies
\begin{equation}
        \delta w \sim 
         {\rm exp} \, [-3/2(w_B-w_0) \ln{a}] \, \delta Q.
\end{equation}
In particular, this equation shows that  $\delta w$  decays 
if $\delta Q$ decays.

To show the $\delta Q$ decays, we
start from the more standard form of the equation-of-motion:
\begin{equation}
\label{eom}
\ddot{Q}+3H\dot{Q}+V'=0,
\end{equation}
 and obtain the perturbed equation
\begin{equation}  \label{eom'}
\delta \ddot{Q} + 3H \delta \dot{Q}
+ V_0''\delta Q=0.
\end{equation}
Changing the variable to $d\tau=Hdt$, we obtain
\begin{equation} 
\frac{d^2\delta Q}{d\tau^2}=\frac{d}{Hdt}\left(\frac{d\delta Q}{Hdt}
\right)=\frac{1}{H^2}\ddot{Q}+m\frac{d\delta Q}{d\tau}
\end{equation}
where  $m\equiv(Ht)^{-1}=3(w_B+1)/2$.
Eq.~(\ref{eom'}) then becomes
\begin{equation}  \label{corr}
\frac{d^2\delta Q}{d\tau^2}+(3-m)\frac{d\delta Q}{d\tau}
+\frac{V_0''}{H^2}\delta Q = 0.
\end{equation}
where 
\begin{equation}
C\equiv \frac{V_0''}{H^2} \approx \frac{9}{4}(1-w_Q)(2+w_B+w_Q)
\end{equation}
in the adiabatic approximation.
The solution to Eq.~(\ref{corr}) is $\delta Q=A\exp{(\beta \tau)}$, where
\begin{equation} \label{beta}
\beta^2+(3-m)\beta +C=0
\end{equation}
with solutions
\begin{equation}
\label{pm}
\beta_{\pm}=-\frac B 2 \left[1\pm \sqrt{1-\frac{4C}{B^2}}\right]
\end{equation}
where
\begin{equation}
\label{B}
B=3-m=\frac 3 2 (1-w_B).
\end{equation}
Note that $B>0$ and $C>0$.
Hence, we have that $\delta Q$ exponentially  decays, 
and,  by the argument that 
preceded, $\delta w\propto \delta Q $ also decays exponentially.
Combining our relations for $\delta Q$ and $\delta w$, we can
reproduce the result in Eq.~(\ref{gam}) obtained for the constant
$\Gamma$ case.

If $Q$ has spatial fluctuations, Eq.~(\ref{corr}) must be modified
by a positive term proportional to $k^2 \delta Q$. 
The effect is to
increase $C$ and modify the oscillation frequency. However, the 
exponential suppression of the fluctuations is retained once the
field starts approaching the attractor solution. Hence, even if
the initial conditions result in
significant fluctuations  after the field is frozen (in either the
undershoot or overshoot case), 
 the initial
fluctuations are erased as $Q$ converges to the tracker solution.



\end{document}